\documentclass[aps,prd,twocolumn,showpacs,superscriptaddress,groupedaddress,nofootinbib]{revtex4-1}
\usepackage[utf8x]{inputenc} 
\usepackage{amsmath}
\usepackage{slashed}
\usepackage[T1]{fontenc}
\usepackage{amstext}
\usepackage{amssymb}
\usepackage{braket}
\usepackage{graphicx}
\usepackage{hyperref}


\begin{document}
 \author{Parsa Hossein Ghorbani}
 \title{\boldmath Electroweak Phase Transition in Scale Invariant Standard Model}
 \affiliation{Institute for Research in Fundamental Sciences (IPM), 
  School of Particles and Accelerators, P.O. Box 19395-5531, Tehran, Iran}
 \email{parsaghorbani@gmail.com}

 \begin{abstract}
In an extension to the scale invariant standard model by two real singlet scalars $s$ and $s'$ in addition to 
the Higgs field, we investigate the strong first-order electroweak phase transition as a requirement for baryogenesis. This
is the minimal extension to the scale invariant standard model with two extra degrees of freedom that possesses
the physical Higgs mass of $125$ GeV.  
The scalar $s'$ being stable because of the $\mathbb Z_2$ discrete symmetry is taken as the dark matter candidate. We then 
show that the electroweak phase transition is strongly first-order, the dark matter relic density takes the
desired value $\Omega_{\text{DM}}h^2 \sim 0.11$, and the constraints from direct detection experiments are respected only if 
$m_{s'}\equiv m_{\text{DM}} \gtrsim 4.5$ TeV. The model also puts a lower bound on the scalon mass, $m_s \gtrsim 200$ GeV.
\end{abstract}

\maketitle 

\section{Introduction}

After the discovery of the Higgs particle in July 2012 at the LHC \cite{Aad:2012tfa,Chatrchyan:2012xdj}, 
the last missing piece of the standard model (SM) prediction, made almost a half-century ago \cite{Englert:1964et,Higgs:1964pj}, 
was completed. The SM has been tested by the 
most stringent scrutinies over many different experiments and it has passed them successfully. 
However there are a number of
issues, either theoretical or experimental/observational, that are not compatible with the SM predictions. 
The gauge hierarchy problem, the strong first-order electroweak phase transition (EWPT)
and other conditions needed in the baryogenesis mechanism, 
and the problem of dark matter are some examples of unanswered puzzles in the SM. These inconsistencies
led people to think of theories beyond the SM such as the GUT, SUSY, etc. Our goal in this paper is to address
the above-mentioned SM shortcomings rather in a minimal extension of the scale invariant standard model (MSISM). 

The negative Higgs mass term, $-m_H^2 H^\dagger H$ in the SM potential causes a quadratical divergent term proportional 
to the energy scale cutoff $\Lambda^2$ after including the quantum corrections. 
In fact, the Higgs mass term is the only term that breaks
the classical scale invariance in the SM. Therefore by omitting the Higgs mass term from the SM potential we have practically 
removed the problem of gauge hierarchy 
\footnote {Even with the vanishing Higgs mass the quadratically
divergent term depends on the regularization scheme and may reappear 
due to the quantum corrections \cite{Meissner:2007xv}. Nevertheless, in dimensional regularization scheme the Higgs remains massless 
to all orders of perturbation\cite{Bardeen:1995kv}.}. 
In the seminal paper of Coleman and E. Weinberg \cite{Coleman:1973jx} 
it is shown 
that in a scale invariant gauge theory the radiative corrections break the scale invariance and that triggers 
the spontaneous symmetry breaking. Following their work 
Gildener and S. Weinberg \cite{Gildener:1976ih} argued that in the SISM the 
radiative corrections break the electroweak symmetry and thereby the Higgs mechanism is restored for the SISM. 
The SISM with only one classically massless scalar (Higgs) cannot be realistic, because as computed by Gildener and S. Weinberg 
from the quantum corrections the Higgs mass can be just as heavy as around $5$ GeV, which is far lighter than the observed Higgs mass, 
$m_H\sim 125$ GeV.  
As discussed in \cite{Gildener:1976ih} in general among the $n$ scalars in addition to the Higgs scalar in the extended SISM, 
there is at least one heavy scalar that may be interpreted as the Higgs particle and there is one classically massless scalar
that is dubbed
 the {\it scalon}. In this paper we add only two extra scalars to the SISM, as the most minimal scale invariant extension of SISM 
that contains the correct Higgs mass. There are papers in the literature 
(see for instance \cite{Farzinnia:2013pga,Sannino:2015wka,Gabrielli:2013hma}) 
that have sought a similar scope but none of them has presented an analytical investigation of the electroweak phase 
transition and furthermore they have usually involved extra fermionic degrees of freedom in the hidden sector. Among papers which address the problem of dark matter using scale invariant extension of the SM with multiple scalars one may refer to Ref. \cite{Endo:2015nba} in which the detectability of the real scalar in direct detection experiments is examined. In for instance Ref. \cite{Hashino:2016rvx}, on the other hand the strongly first-order electroweak phase transition is studied in the scale invariant SM with additional isospin singlet scalar fields.
In the current work, we are interested only in studying 
the pure scalar and minimal extension of the SISM, investigating both the DM and the EWPT at the same time.

This paper is organized as the following. In the next section we build up the model extending the SISM with two real scalars. 
Then in section \ref{wash} we derive the critical temperature and the washout criterion for the electroweak phase transition.
In section \ref{stability} the stability conditions are given and the next section will be on dark matter computations. 
Finally in section 
\ref{dir} we examine the model with the experimental bounds on dark matter elastic scattering off the nucleus. 
We conclude in section \ref{conc}.

\section{Minimal Extension of Scale Invariant Standard Model}

In the SM, if we set the Higgs mass term to zero, the only term remaining in the Higgs potential will be  
$\lambda (H^\dagger H)^2$ or  $\lambda h^4$ after gauging away three components of the Higgs doublet. 
In Gildener-Weinberg notation \cite{Gildener:1976ih}, the scalar potential
with $n$ scalars, $s_i$,  is shown as $1/24 \lambda_{ijkl}\,s_i s_j s_k s_l$ where $\lambda_{ijkl}$ 
denotes the coupling. 
As discussed in \cite{Ghorbani:2015xvz}, in order to have a scale invariant version of the standard model 
possessing a Higgs doublet with 
the observed Higgs mass of $125$ GeV as well as other SM particles with their 
physical masses, at least two more scalars (singlet) must be added to the theory. The reason comes from 
eq. (4.6) of \cite{Gildener:1976ih} where the scalon gaining mass through the radiative corrections, depends 
only on the masses of the Higgs particle, the gauge bosons and the mass of the top quark. In the absence of any additional 
scalar except the Higgs and the scalon, this expression will be negative.
In this paper, therefore we stay in the most minimal potential with only two extra scalars which we call them here $s$ and $s'$. 
We also assume that 
these scalars appear with the $\mathbb{Z}_2$ symmetry in the potential to attribute them later to the dark matter candidate, 
\begin{equation}
\begin{split}
V_{\text{tr}}(h,s,s') & =\frac{1}{4}\lambda_{h}h^{4}+\frac{1}{2}\lambda_{hs}s^{2}h^{2} \\
& +\frac{1}{4}\lambda_{s}s^{4}+\frac{1}{2}\lambda_{ss'}s^{2}s'^{2}+\frac{1}{4}\lambda_{s'}s'^{4}\,,\label{lag1}
\end{split}
\end{equation}
where $h$ stands for the Higgs field, $H^\dagger=\frac{1}{\sqrt{2}} (0~ h)$. At high temperature, above 
 the electroweak phase transition temperature,
the theory lives in its symmetric phase and the vacuum expectation values ({\it vev}) of the
fields are temperature dependent. Let us assign the {\it vev} of each field as 
\begin{equation}
v_h\left(T\right)\equiv \braket{h} ,\,\,\,\,\,v_s\left(T\right)\equiv\braket{s},\,\,\,\,\,\,v_{s'}\left(T\right)\equiv\braket{s'}.
\end{equation}
We require that the vacuum expectation values after the phase transition
be $(v_{h}=246$ GeV$, v_{s}\neq 0, v_{s'}=0)$ so that the scalar
field $s'$ remains stable because of the $\mathbb{Z}_{2}$ symmetry
and thus it can play the role of the dark matter candidate.  We see latter that the value of $v_{s}$ is not fixed 
and it depends on the value of the couplings in the model. 

In the Lagrangian in eq. (\ref{lag1}), the term $\lambda_{hs'} h^2 s'^2$  could in principle be considered  because $s'$ undergoes no {\it vev}. Although this term contributes in the DM relic density, we know from the singlet Higgs portal model (SHP) that $\lambda_{hs'}$ must be small to evade the direct detection bounds. It is therefore reasonable to assume that vanishing. The inclusion of the $\mathbb{Z}_2$ odd terms $s's^3$ and $s'^3 s$ in the Lagrangian is allowed by the scale invariance but will lead to the decay of the DM scalar $s'$ and the observed relic density will not be obtained, hence we avoided such terms in the Lagrangian.

In the scale invariant standard model the flat direction is defined as the direction along which the tree-level potential is vanishing. 
This condition is equivalent 
to imposing the Ward identity of the scale symmetry in a scalar theory \cite{AlexanderNunneley:2010nw}. 
The flat direction for the potential in eq. (\ref{lag1}) is obtained via a rotation in the $\left(h,s\right)$
space by the angle $\alpha$, 
\begin{equation}
\cos^{2}\alpha=- \frac{\lambda_{hs}}{\lambda_{h}-\lambda_{hs}},\,\,\,\,\lambda_{hs}^{2}-\lambda_{h}\lambda_{s}=0.\label{flat-con}
\end{equation}
The mass matrix (being meaningful after
the phase transition at low temperature) is off-diagonal
only in $\left(h,s\right)$ block. This is because of our special
choice; we have taken a nonzero {\it vev} for $h$ and $s$ and vanishing
{\it vev} for $s'$. Finally, the mass eigenvalues after the EWPT
read, 
\begin{equation}
m_{h}^{2}=2v_{h}^{2}(\lambda_{h}-\lambda_{hs}),\,\,\,\,m_{s}^{2}=0,\,\,\,\,\,m_{s'}^{2}=
-v_{h}^{2}\frac{\lambda_{h}\lambda_{ss'}}{\lambda_{hs}}.\label{mass}
\end{equation}

The one-loop correction at zero temperature gives a small mass to
the classically massless eigenstate $s$, the so-called {\it scalon}
field \cite{Gildener:1976ih}, 
\begin{equation}
\delta m_{s}^{2}=\frac{-\lambda_{hs}}{32\pi^{2}m_{h}^{2}}\left(m_{h}^{4}
+m_{s'}^{4}+6m_{W}^{4}+3m_{Z}^{4}-12m_{t}^{4}\right)\label{scalon-mass}\, .
\end{equation}
Without the introduction of the second singlet scalar $s'$ while having the observed Higgs mass of $125$ GeV
and the correct masses for the top quark and gauge bosons, 
the mass correction to the scalon field 
$s$ could not be positive. 
Now by means of the radiative correction, the scalon can be interpreted as the mediator in DM models. This can be seen from eq. (\ref{scalon-mass}). For more details see \cite{Ghorbani:2015xvz}.

The one-loop effective potential consists of the tree-level potential
in eq. (\ref{lag1}), the Coleman-Weinberg one-loop correction at
zero temperature, and the one-loop correction at finite temperature,
\begin{equation}
V_{\text{eff}}=V_{\text{tr}}+V_{0}^{\text{1-loop}}+V_{T}^{\text{1-loop}}.\label{veff}
\end{equation}
If $T\gg m_{i}$ with $m_{i}$ the tree-level mass of particle $i$,
the one-loop thermal contribution is approximated as $V_{T}^{\text{1-loop}}\simeq CT^{2}\phi^{2}$
and the total one-loop effective potential becomes \cite{Marzola:2017jzl}, 
\begin{equation}
V_{\text{eff}}\simeq-\frac{1}{4}B\phi^{4}+\frac{1}{2} B\phi^{4}\log\frac{\phi^{2}}{v_{\phi}^{2}}+CT^{2}\phi^{2} \,,\label{v1-loop}
\end{equation}
where $\phi$ is the radial field in the polar coordinate system, 
i.e. $\left(h,s\right)\equiv\left(\phi\cos\alpha,\phi\sin\alpha\right)$\footnote{Note that 
we are considering a one-step phase transition along the flat direction $\phi$. In general, if more scalars in the model take non-zero {\it vev} a two-step phase transition may be required. See 
\cite{Chao:2017vrq} for instance},
hence $v_{\phi}^{2}=v_{h}^{2}+v_{s}^{2}$. The coefficients $B$ and
$C$ are, 
\begin{equation}\label{B}
B=\frac{1}{64\pi^{2}v_{\phi}^{4}}\left(m_{h}^{4}+m_{s'}^{4}+6m_{W}^{4}+3m_{Z}^{4}-12m_{t}^{4}\right)
\end{equation}
\begin{equation}\label{C}
C=\frac{1}{12v_{\phi}^{2}}\left(m_{h}^{2}+m_{s'}^{2}+6m_{W}^{2}+3m_{Z}^{2}+6m_{t}^{2}\right).
\end{equation}
In case of no mixing, i.e. when $\cos\alpha=0$ then $v_\phi=v_H$ and the problem turns into studying the electroweak symmetry breaking in the scale invariant standard model which has been investigated in \cite{Endo:2015ifa,Hashino:2015nxa}.
\section{Critical Temperature and Washout Criterion }\label{wash}

The strong first-order electroweak phase transition is one of the three Sakharov conditions \cite{Sakharov:1967dj} for the baryogenesis. 
For the {\it CP} violation in the minimal scale invariant extensions of the SM see  \cite{AlexanderNunneley:2010nw}.
The phase transition takes place at the critical temperature, $T_c$ at which
the free energy (effective potential) has two degenerate minima at $T=T_{c}$. In this section we follow \cite{Ghorbani:2017jls} to calculate 
analytically the washout criterion i.e. $v(T_c)/T_c>1$ which guarantees 
the strong first-order phase transition.  

The minimization condition on the thermal effective potential in
eq. (\ref{v1-loop}) with the derivative being along the radial field, 
\begin{equation}
\frac{\partial}{\partial\phi}V_{\text{eff}}\Bigl|_{v_\phi\left(T\right)}=0,
\end{equation}
leads to a set of $T$-dependent equations for the vacuum expectation
value, 
\begin{equation}
v_{\text{sym}}(T)=0,
\end{equation}
\begin{equation}
v_{\text{brk}}^{2}(T)\log\frac{v_{\text{brk}}^{2}(T)}{v_{\phi}^{2}}=-\frac{C}{B}T^{2}\,,\label{vbrk}
\end{equation}
where $v_{\text{sym}}$ is the {\it vev} of the radial field in the symmetric phase and $v_{\text{brk}}$ is the {\it vev} in the broken 
phase. In the SM, the $v_{\text{brk}}$ is the temperature-dependent vacuum expectation value of the Higgs doublet, but in our case $v_{\text{brk}}$ is the {\it vev} along the flat direction, i.e. the {\it vev} of the mixing of the Higgs doublet and the scalon.
Eq. (\ref{vbrk}) has no analytic solution for $v_{\text{brk}}$.
Nevertheless, the solution can be expressed in terms of the {\it Lambert W function}
which is defined as, 
\begin{equation}
z=we^{w}\Leftrightarrow w=W\left(z\right),
\end{equation}
where $z$ and $w$ in general are complex numbers. In terms of the
Lambert W function eq. (\ref{vbrk}) is written as, 
\begin{equation}
v_{\text{brk}}^{2}=\frac{-CT^{2}/B}{W\left(-\frac{C}{Bv_{\phi}^{2}}T^{2}\right)}.\label{Lam1}
\end{equation}

\begin{figure}
\centering
\includegraphics[scale=0.45]{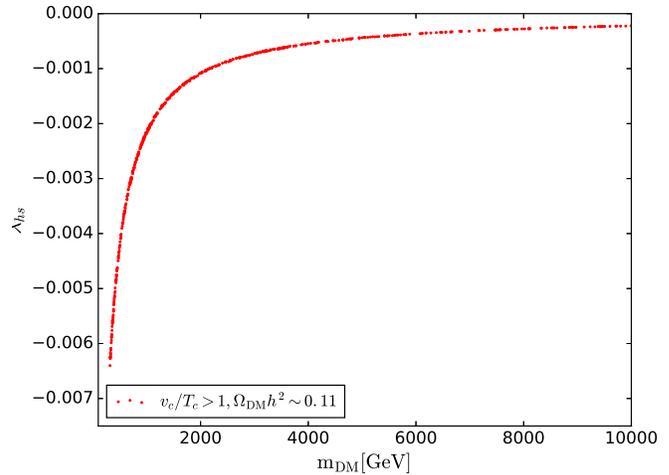}
\caption{The plot compares the dark matter mass against the coupling $\lambda_{hs}$ with washout criterion 
satisfied and $\Omega_{\text{DM}}h^2 \sim 0.11$.}
\label{lhs}
\end{figure}

At the critical temperature $T_{c}$ the effective potential in eq. (\ref{v1-loop}) must be vanishing at the
minimum $v_{\text{brk}}$ as it is vanishing also at the minimum $v_{\text{sym}}=0$ of the symmetric phase. 
Multiplying eq. (\ref{vbrk}) by $v_\phi^2$ and substituting its right-hand side in $V_{\text{eff}}(v_{\text{brk}})=0$ from
eq. (\ref{v1-loop}) we arrive at  $v_{\text{brk}}^2(T_c)=(2C/B) T_c^2$. Therefore the condition for the electroweak 
phase transition to be strongly first-order (the washout criterion) becomes,

\begin{equation}\label{vbrk1}
 \frac{v_{\text{brk}}(T_c)}{T_c}=\sqrt{\frac{2C}{B}} > 1\,.
\end{equation}

Finally,  substituting $v_{\text{brk}}$ from eq. (\ref{vbrk1}) into eq. (\ref{v1-loop}), expanding the effective potential
 and setting that to zero we obtain, 
\begin{equation}\label{Tc2}
T_{c}^{2}\simeq\left(\sqrt{11}-3\right)\frac{B}{C}v_{\phi}^{2}\,.
\end{equation}

Before going further with more constraints on the parameters, regarding the values of $B$ and $C$ in 
eqs. (\ref{B}) and (\ref{C}), it is clear that the ratio $v_c/T_c$ can easily be large enough leading to a very strong first-order
phase transition.

\section{Stability Conditions}\label{stability}

The stability conditions impose already strong constraints on the
parameters of the model. The first derivative of the tree-level potential
in eq. (\ref{lag1}) must vanish at the {\it vev}s, 
\begin{equation}
\frac{\partial V}{\partial h}\Bigl|_{\braket{h}}=\frac{\partial V}{\partial s}\Bigl|_{\braket{s}}=0\,,\\
\end{equation}
which in turn leads to, 
\begin{equation}
 \lambda_{h}v_{h}^{2}=-\lambda_{hs}v_{s}^{2}\,,\hspace{1em}\hspace{1em}\lambda_{s}v_{s}^{2}=
 -\lambda_{hs}v_{h}^{2}\,.\label{dV}
\end{equation}

The positivity of the second derivatives of the potential in eq. (\ref{lag1}) gives rise to, 
\begin{equation}
\lambda_{hs}<0,\hspace*{1em}\lambda_{ss'}>0.\label{stab1}
\end{equation}
From eq. (\ref{stab1}) and eq. (\ref{mass}) we get $\lambda_{h}>0.$
Now the radiative correction to scalon mass in eq. (\ref{scalon-mass})
is positive if $m_{s'}>316.5$ GeV. Using eq. (\ref{mass}) for $m_{s'}$
one arrives at $\lambda_{ss'}>-1.65\lambda_{hs}/\lambda_{h}.$ Still
we can make use of the Higgs mass relation in eq. (\ref{mass}) to
constrain more the Higgs coupling: $\lambda_{h}=\lambda_{hs}+0.128$.
As $\lambda_{h}>0$ and $\lambda_{hs}<0$ then $-0.128<\lambda_{hs}<0$.

\begin{figure}
\centering
\includegraphics[scale=0.45]{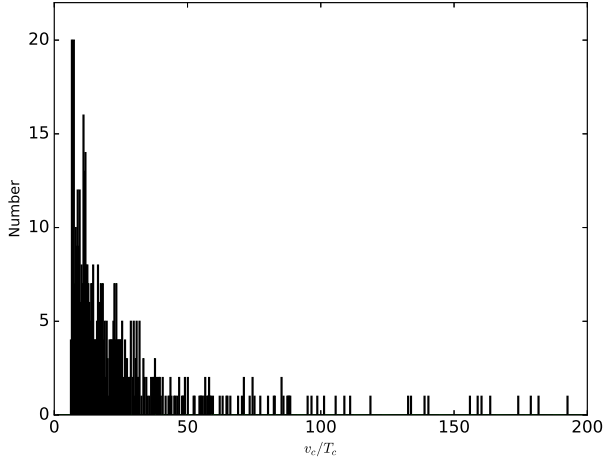}
\caption{A histogram for the ratio $v_c/T_c$ with the correct relic density $\Omega h^2\sim 0.11$. 
It is shown that $v_c/T_c \gtrsim 4$ which guarantees a very strong first-order phase transition. }
\label{histo}
\end{figure}

\section{Dark Matter}\label{dm}

The scalar $s'$ taking a zero expectation value is stable and can play the role of the thermal dark matter candidate within the 
freeze-out scenario. In this section 
we add the relic density condition to the washout criterion obtained in the previous section and probe the space of the parameters. 
The independent parameters in the model are not many; $\lambda_{hs},\lambda_{s'}$ and $ \lambda_{ss'}$ among which only 
the parameter $\lambda_{hs}$ takes part in the relic density computation. The dark matter sector interacts with the visible sector via 
the scalar mediator $s$ which has become massive through the radiative correction and its mass is given by eq. (\ref{scalon-mass}). 
In fact the scalar mediator $s$ mixes with the Higgs field in the SM and the mixing angle is that of the flat direction in 
eq. (\ref{flat-con}). The Higgs vacuum expectation value is known experimentally; $v_h=246$ GeV and the {\it vev} 
of the scalar $s$ is determined by other known parameters of the theory as seen from eq. (\ref{dV}).

\begin{figure}
\centering
 \includegraphics[angle=-90,scale=0.35]{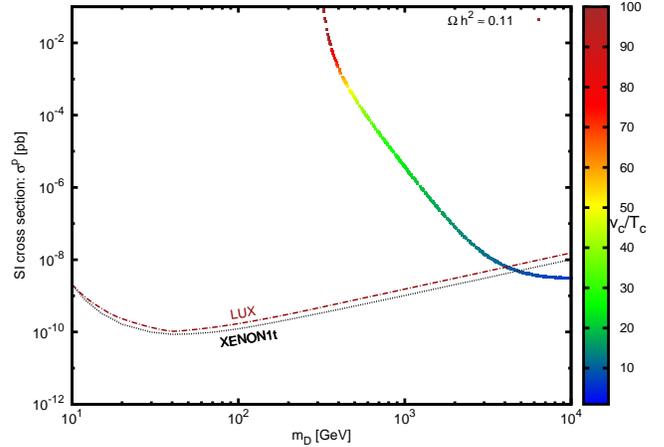}
\caption{The viable range of the DM mass is $m_{\text{DM}}\gtrsim 4.5$ TeV after imposing the XENON1t/LUX direct detection 
experiments on DM-nucleus elastic 
scattering cross section, the DM relic density constraint, and the first-order phase transition condition. }
\label{direct}
\end{figure}

The thermal evolution of the dark matter number density, $n_{s'}$ in the early Universe is given by the Boltzmann equation,
\begin{equation}\label{boltz}
\frac{dn_{s'}}{dt}+3Hn_{s'}= -\braket{\sigma_{\text{ann}} 
v_{\text{rel}}} \left[ n_{s'}^2-\left( n^{\text{EQ}}_{s'}\right )^2 \right] \,,
\end{equation}
where $H$ is the Hubble expansion rate, $v_{\text{rel}}$ stands for the dark matter relative velocity and  
the $\sigma_{\text{ann}}$ is the dark matter annihilation cross section. 
We compute the relic abundance using the {\tt MicrOMEGAs4.3} package 
\cite{Belanger:2014vza} that numerically solves the Boltzmann differential equation.
We recall that the potential we use to compute the relic density is the potential in eq. (\ref{lag1}) after the 
electroweak symmetry breaking which is given by,  
\begin{equation}\label{potential}
\begin{split}
 & V(h,s,s')= \frac{1}{2}m_h^2 h^2 +\frac{1}{2} m_{s'}^2 s'^2\\
 &+ \left( \lambda_h + \lambda_{hs} \right) \sqrt{1-\frac{\lambda_{hs}}{\lambda_h}} v_h h^3
 + \frac{1}{4} \frac{\left( \lambda_h + \lambda_{hs} \right)^2}{\lambda_h} h^4 \\
 &+ \left( \lambda_h  + \lambda_{hs} \right) \sqrt{-\frac{\lambda_{hs}}{\lambda_h}} h^3 s \\
 &+ 2\sqrt{-\lambda_{hs} \left (\lambda_h -\lambda_{hs} \right) } v_h h^2 s\\
 &- \lambda_{hs} h^2 s^2 
 + \frac{\lambda_h \lambda_{ss'} v_h}{\sqrt{-\lambda_{hs} (\lambda_h-\lambda_{hs})}} s s'^2\\
& - \frac{\lambda_{ss'} \lambda_h v_h}{\sqrt{\lambda_h (\lambda_h-\lambda_{hs})}} h s'^2
 + \frac{\sqrt{-\lambda_{hs} \lambda_h}\lambda_{ss'}}{\lambda_h-\lambda_{hs}} s h s'^2 \\
 &+\frac{1}{2} \frac{\lambda_h \lambda_{ss'}}{\lambda_h-\lambda_{hs}} s^2 s'^2
 -\frac{1}{2} \frac{\lambda_{ss'} \lambda_{hs}}{\lambda_h-\lambda_{hs}} h^2 s'^2
 +\frac{1}{4} \lambda_{s'} s'^4\,.
 \end{split}
\end{equation}

Note that the Higgs scalar field has a mass term now in eq. (\ref{potential}). The phase transition  
(going from the symmetric phase with $v_h=0$ to the broken phase with $v_h \neq 0$) is followed by
the scale symmetry breaking through the radiative correction to the scalon mass. 
We constrain the model by the observed dark matter relic abundance from the WMAP/Planck \cite{Hinshaw:2012aka,Adam:2015rua}
to be $\Omega_{\text{DM}}h^2 \sim 0.11$. In section \ref{stability} the mass of the dark matter already had  a lower bound
due to the positivity of the scalon mass; $m_s'\equiv m_{\text{DM}}>316.5$ GeV. In Fig. \ref{lhs} the 
dark matter mass is plotted against the only independent coupling i.e. the $\lambda_{hs}$. As seen from 
this figure the viable range of the coupling shrinks into $ -0.007 \lesssim \lambda_{hs} \lesssim 0$. The scalon could be searched at the Large Hadron Collider (LHC) or future colliders via the exotic Higgs decays $h\to ss$ and $h\to sss$. It has been pointed out in Ref. \cite{Curtin:2013fra} that even couplings as small as $\mathcal{O}(10^{-2})$ yield Br($h\to$BSM)$\sim 10\%$. So even very small Higgs-scalon coupling 
can in principle lead to a signature at the LHC.
The DM 
mass however sits almost within the same limit we obtained in section \ref{stability} i.e. $m_{\text{DM}}>318.3$ GeV. 
In Fig. \ref{histo} we have also demonstrated a histogram of the values $v_c/T_c$ which are bounded by the 
correct relic density. It is understood that interestingly $v_c/T_c$ is greater than $3.8$ and much 
bigger that guarantees a very strong first-order 
electroweak phase transition. 

\section{Direct Detection Constraint}\label{dir}
There are experiments that have been set up with the goal of detecting the elusive dark matter directly.
Among these, the XENON1t experiment located at Gran Sasso in Italy is the most recent and the 
more accurate one \cite{Aprile:2017iyp}. 
Although the XENON1t experiment and no other experiments such as LUX (see \cite{Akerib:2016vxi} for the recent results), 
have not detected the dark matter but they have put a very stringent constraint on the 
elastic scattering cross section of the dark matter off the nucleus. 
We examine the current model by data from the direct detection experiments. 
\begin{figure}
\centering
 \includegraphics[scale=0.45]{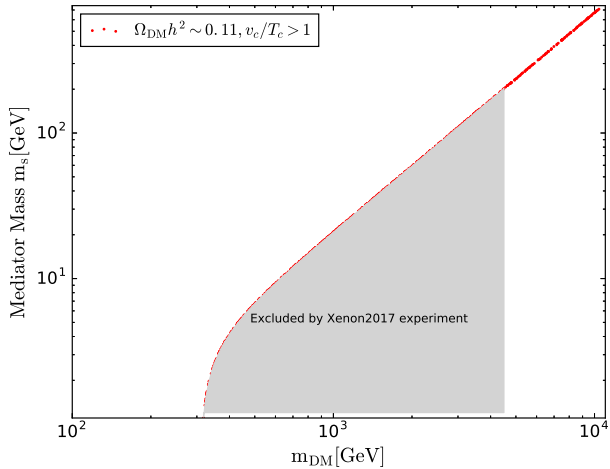}
\caption{The plot shows the viable range of the scalon mass being $m_s\gtrsim 200$ GeV. }
\label{ms}
\end{figure}

The DM-nucleus cross section can be described simply by the following effective potential,
\begin{equation}
\mathcal L_{\text{eff}} = \alpha_{q} s' s' ~ \bar q q \,,   
\end{equation}
where $q$ stands for the quark in the nucleon and $s'$ is the dark matter field. The coupling  $\alpha_{q}$ is given by, 
\begin{equation}
\alpha_{q} = m_{q} \frac{2\lambda_h \lambda_{ss'}}{\lambda_h -\lambda_{hs}}  (\frac{1}{m_{h}^2} - \frac{1}{m_{s}^2} )\,.
\end{equation}
The DM-nucleus scattering is obtained from
a tree-level Feynman diagram leading to the following spin-independent elastic scattering cross section, 
\begin{equation}\label{Ncross}
\sigma^{\text{N}}_{\text{SI}} = 
\frac{\alpha_{N}^2  \mu_{N}^2}{\pi m_{\text{DM}}^2}, 
\end{equation}
where $\mu_{N}$ is the reduced mass for the DM-nucleus system and $\alpha_{N}$ denotes a coefficient 
that depends on the nucleon form factors. For more details on $\alpha_{N}$ see \cite{Ghorbani:2014qpa} and the references therein. 

For the viable parameter space that we obtained in section \ref{dm} we have computed the elastic scattering 
cross section in eq. (\ref{Ncross}) using the {\tt MicrOMEGAs4.3} package. The result is shown in Fig. \ref{direct}.
According to Fig.  \ref{direct} only the dark matter mass $m_{\text{DM}}\gtrsim 4.5$ TeV survives the XENON1t/LUX cross section limits 
\footnote {Thanks to Christopher Tunnell for providing me with the XENON1t data in Fig. 4 of \cite{Aprile:2017iyp}.}
while 
respecting both the relic density constraint and the washout criterion.
It is interesting also to determine the allowed masses of the scalon $s$ which is 
the mediator connecting the DM sector to the SM. As seen in Fig. \ref{ms} the scalon mass $m_s$, takes only values above $200$ GeV.  

\section{Conclusion}\label{conc}
In this paper we  have studied the minimal extension of the scale invariant standard model with two extra scalars, 
$s$ and $s'$ in 
addition to the Higgs particle. Two scalars are the minimum number of scalars 
we should add to the scale invariant standard model to give a mass of $125$ GeV to the Higgs and 
correct masses for other particles in the SM. 
The classically massive scalar, $s'$ is interpreted as a dark matter candidate and the classically massless scalar, $s$
called the scalon plays the role of the DM-SM mediator.  
We showed that this model supports a very strong first-order electroweak phase transition even 
if we constrain the model with the observed DM relic density by WMAP/Planck. Imposing the limits from the direct detection experiments 
such as XENON1t/LUX on the elastic scattering cross section of the DM-nucleus still allows the  dark matter mass $m_{\text{DM}}\gtrsim 4.5$ TeV
and the scalon mass $m_s \gtrsim 200$ GeV.\\

\begin{acknowledgments}
I am very grateful to Karim Ghorbani for many fruitful discussions.  I would also like to thank Ville Vaskonen for useful email
correspondence. 
\end{acknowledgments}

\bibliographystyle{apsrev4-1}
\bibliography{ref.bib}

\end{document}